\documentstyle[12pt]{article}

\def\cmm2{{\,\rm cm^{-2}}}
\def\cm2{{\,{\rm cm}^2}}
\def\cmm3{{\,{\rm cm}^{-3}}}
\def\gcmm3{{\,{\rm g\,cm^{-3}}}}

\def\gtwid{\mathrel{\raise.3ex\hbox{$>$\kern-.75em\lower1ex\hbox{$\sim$}}}}
\def\ltwid{\mathrel{\raise.3ex\hbox{$<$\kern-.75em\lower1ex\hbox{$\sim$}}}}

\def\fun#1#2{\lower3.6pt\vbox{\baselineskip0pt\lineskip.9pt
  \ialign{$\mathsurround=0pt#1\hfil##\hfil$\crcr#2\crcr\sim\crcr}}}

\begin{document}
\thispagestyle{empty}

\begin{center}
\rightline{CfPA-94-Th-22}
\rightline{astro-ph/9404???}
\rightline{Submitted to ApJ }

\vspace{0.3in}
{\Large\bf Sample Variance of  non-Gaussian  Sky Distributions}\\

\vspace{.3in}
{\large\bf Xiaochun Luo  } \\
\vspace{0.2in}

{ Center for Particle Astrophysics\\
301 Le Conte Hall\\
  University of California, Berkeley, CA  94720}\\

\end{center}

\vspace{.3in}
\centerline{\bf ABSTRACT}
\vspace{0.3in}
Non-Gaussian distributions of cosmic microwave background (CMB) anisotropies
have been proposed to  reconcile the discrepancies  between different
experiments  at half-degree scales (Coulson et al. 1994).  Each experiment
probes a
different part of the sky, furthermore,
sky coverage  is very small, hence the sample variance
of each experiment can be large, especially when the sky signal
is non-Gaussian.  We model the degree-scale CMB sky as a
$\chi_{n}^{2}$ field with $n$-degrees of freedom and show that
the sample variance is enhanced over that of a
Gaussian  distribution by a factor
of $ {(n +6)/ n}$. The sample variance for different experiments are
 calculated, both for Gaussian and non-Gaussian distributions.
  We  also show that
if the distribution is highly
non-Gaussian $(n \ltwid 4)$  at half-degree scales,
 then  the non-Gaussian signature of the CMB could be  detected
 in the FIRS map, though
probably not in the COBE map.

\centerline{ {\it Subject headings:} cosmic microwave background -
cosmology: theory}
\newpage

Recently, several groups have reported results on the measurement
of anisotropies of the cosmic microwave background (CMB) at
degree scales (de Bernardis et al. 1994; Cheng et al. 1993; Gunderson et al.
1993; Meinhold et. al 1993; Schuster et al. 1993; Tucker et al. 1993;
Wollack et al. 1993). The beam size, beam throw,
 most sensitive angular scale,
 sky coverage and quoted $\rm rms$ temperature anisotropies are
summarized in Table 1. The results from these experiments do not agree with
each other, in particular the results from the same experiment,  MAX, for
two different part of the sky, the Gamma Ursa Minor (GUM) region and the
mu-Pegasi  (MuP) region,   contradict each other at $2 \sigma$ level.
 A way to reconcile these measurements is to have
a non-Gaussian distribution of temperature anisotropies at half-degree
scales (e.g. Coulson et al. 1994).
At present, there are still large uncertainties in all experiments due to
 foreground subtractions, therefore the need to invoke
 non-Gaussian temperature distributions
remains to be established.
Since different experiments probe different
part of the sky and the sky coverage of each experiment is small, the
 sample variance of each experiment can be  large, especially
when the sky signal is non-Gaussian.
The goal of this paper is to quantify the difference in the
expected sample variance between non-Gaussian and Gaussian fields
in order to determine if this effect could be responsible for
the discrepancy between experiments. We also
estimate the minimum sample size for each experiment
in order for the sample variance to be less than $20
\mu K$, both for Gaussian and non-Gaussian distributions.

Regarding the  statistical analysis, the most important
quantity is the number of independent measurements in a single
sample.
However, the  data points in CMB experiments are correlated
and therefore contain less statistical information.
Thus,  it is useful to determine the  effective number of data points,
 $N_{e}$,  defined as the number of independent
measurements of temperature anisotropies in each experiment
(we will discuss how to estimate $N_{e}$ for each experiment later).
 Expressed
in terms of $N_{e}$,
the sample-averaged {\it rms} temperature anisotropy is
\begin{equation}
\biggl({\delta T\over T}\biggr)_{sam}^{2} =  \Delta  = {1\over N_{e}}
\sum_{i =1}^{N_{e}} \biggl({\Delta T_{i}\over T}\biggr)^{2},
\end{equation}
where   ${\Delta T_{i}/ T}$ is an independent measurement of
 temperature anisotropy at an angular scale corresponding to the beam size.
 The ensemble average
of   ${\Delta T_{i}/ T}$ vanishes and the ensemble average  of
  $\Delta$ is $\sigma_{th}^{2}$, where $\sigma_{th}$ is
 the theoretical prediction for the temperature anisotropy at the angular
scale probed by the experiment.

Due to the stochastic nature of ${\Delta T \over T}$,
$ \Delta$ will   be a random quantity which can vary between
different  experiments.
The sample variance is the  fluctuation around $\sigma_{th}^{2}$,
the ensemble-average of $\Delta$.
In the case where ${\Delta T \over T}$ is  Gaussian,  the sample variance
is given by
\begin{equation}
\sigma_{sam}^{2} = \langle \Delta^{2}\rangle
- \langle \Delta \rangle^{2} = {2\over N_{e}} \langle \Delta \rangle^{2},
\  \ \langle \Delta \rangle = \sigma_{th}^{2}.
\end{equation}
Here the angle-brackets $\langle ... \rangle$ denote an ensemble average.
 The sample
variance will be reduced by increasing  the sky coverage by an amount which
will  scale with
the effective number of data points $N_{e}$ as $\sigma_{sam} \propto {1/
\sqrt{N_{e}}}$.

We would expect the sample variance to be larger when the
temperature anisotropies are non-Gaussianly distributed.
 Since the functional space of non-Gaussian
distributions is unlimited,  one is faced with the  question
of choosing  appropriate distributions to model the sky.
As we have stressed  before (Luo 1994), a $\chi_{n}^{2}$
distribution with n-degrees of freedom is one of the simplest and
most natural choices. By varying $n$, it provides a family of distributions
that range from highly non-Gaussian (small $n$) to nearly Gaussian (large $n$).
Furthermore,
it  provides a good fit to the statistics of temperature
fluctuations from global topological defects and non-topological defects in
the framework of the $O(N)   \sigma$-model, where
 a global symmetry $O(N)$ is broken to $O(N-1)$ by an N-component
real scalar field $\phi = (\phi_{1}, ..., \phi_{N})$ in the early universe
(Turok \& Spergel 1991).
In this paper, we will model the degree scale CMB sky as a
$\chi_{n}^{2}$ field,
with
\begin{equation}
{\Delta T \over T} = \sum_{j=1}^{n} (\delta_{i}^{2} - \sigma^{2}),
\end{equation}
where $\delta_{i}, i=1...n$, are $n$-independent Gaussian variables
 with zero mean
and variance $\sigma^{2}$. The ensemble average of $\Delta$ is related to
$\sigma$ through $\langle \Delta \rangle = 2 n \sigma^{4}$.
To calculate the sample variance of the $\it rms$ temperature anisotropy of a
$\chi^{2}$ field, we will utilise some of the higher moments of a Gaussian,
i.e.,
\begin{equation}
\langle \delta^{4}\rangle = 3 \langle\delta^{2}\rangle^{2}, \ \
\langle \delta^{6}\rangle = 15 \langle\delta^{2}\rangle^{3}, \ \
\langle \delta^{8}\rangle = 105 \langle\delta^{2}\rangle^{4},
\end{equation}
and the following identities for Gaussian variables:
\begin{equation}
\langle \sum_{ij=1}^{n} \delta_{i}^{2} \delta_{j}^{2}\rangle = n (n +2)
\sigma^{4},
\end{equation}
\begin{equation}
\langle \sum_{ijk=1}^{n} \delta_{i}^{2} \delta_{j}^{2} \delta_{k}^{2} \rangle
 = n (n +2) (n +4) \sigma^{6},
\end{equation}
and
\begin{equation}
\langle \sum_{ijkl=1}^{n} \delta_{i}^{2} \delta_{j}^{2} \delta_{k}^{2}
\delta_{l}^{2} \rangle = n (n +2) (n +4) (n +6) \sigma^{8}.
\end{equation}
 After some algebra we have
\begin{equation}
\sigma_{sam}^{2} = \biggl({2\over N_{e}}\biggr)\cdot {  n + 6 \over n}  \langle
({\delta T\over T})_{sam}\rangle^{2}
\end{equation}
This is the main result of the paper. For a $\chi^{2}_{n}$ distribution, the
sample variance is enhanced
 by a factor of $ { ( n + 6)/  n} $ relative to a Gaussian distribution.
As we expected,  as n becomes much larger than 6,
 the enhancement is negligible.

The effective number of data points,  $N_{e}$, depends
on the detailed sampling scheme. If the experimental data are sparsely
sampled, i.e. the distance between data points is much larger than the
beam size,
 then $N_{e}$
is approximately the number of experimental data points. If the data is
over-sampled so that correlations between the data points are important, then
$N_{e}$ can  be estimated as  follows.
The sample variance, $\sigma_{sam}$, of a Gaussian-distributed temperature
 anisotropy can also be expressed as (Scott et al 1994):
\begin{equation}
\sigma_{sam} = { 2 \over \Omega^{2}} \int d\Omega_{1} d \Omega_{2}
C^{2}( \theta_{1}, \theta_{2}),
\end{equation}
here $\Omega$ is the solid angle covered by the experiment and $C( \theta_{1},
\theta_{2})$ is the two-point temperature correlation
function. Combining equations (9) and (2) gives an expression for
$N_{e}$:
\begin{equation}
N_{e} = {\Omega^{2} C^{2}(0) \over \int d\Omega_{1} d \Omega_{2}
C^{2}( \theta_{1}, \theta_{2})}.
\label{sample}
\end{equation}
 The data analysis of all experiments involves using the
Gaussian auto-correlation function (GACF), where the correlation among the data
points is approximated as
\begin{equation}
C(\theta) = C(0) \exp(-{\theta^{2}/ 2\theta_{c}^{2}}),
\end{equation}
with $\theta$ the angular distance between two data points and $\theta_{c}$
 the coherence angle.
Since the two point correlation function depends only  on the difference
between two angular directions, after changing variable equation
(\ref{sample}) can be reduced to
\begin{equation}
N_{e} =    \Omega C^{2}(0) /\int d\Omega
C^{2}( \theta).
\end{equation}
For a single one dimensional temperature scan, it
 is easy to evaluate $N_{e}$ since
\begin{equation}
{1\over \Omega C^{2}(0) } \int  d\Omega C^{2} (\theta) = {\sqrt{\pi}}
{\theta_{m} \over \theta_{c}}, \ \ {\rm for} \ \  \theta_{m} \gg \theta_{c},
\end{equation}
where $\theta_{m}$ is the maximum angular difference among data points
with respect to the observer. The solid angle
 covered by the experiment is $\Omega = \theta_{m} \times \theta_{FWHM}$,
hence one can express $N_{e}$ in terms
of $\Omega$ as
\begin{equation}
N_{e} = {1 \over \sqrt{\pi}} ({\Omega \over\theta_{FWHM} \theta_{c}} ).
\end{equation}
Although the above formula is derived for a single one dimensional scan, it is
also applicable to  multiple scan experiments
when  the correlation among different scans is small.  Estimates
of $N_{e}$ for the different experiments  are listed in Table 2.

The sample variance, $\sigma_{sam}$, is directly proportional to the
theoretical prediction of the temperature anisotropy. The  theoretical
model which we  used to calculate the numerical value of $\sigma_{sam}$ is the
standard CDM  ($\Omega =1$) model,  with a scale invariant primordial
power spectrum. We take the  Hubble constant to be  $h =0.5$
and the baryonic faction  $\Omega_{b} = 0.06$, for
consistency with the big-bang nucleosynthesis bound (Walker et al.  1991).
 Theoretical values of the  temperature anisotropy
at degree scales for the experiments
are taken from White {\it et al.} (1994).

 A question of practical interest is just  how large
should  a sample  be in order for the  experimental result to be
informative?   One  criteria could be that the sample variance is much smaller
 the the instrumental sensitivity in $\Delta T$. From equations (2) and (8),
 we estimate
 the minimum sample size
$N_{min}$ that gives rise to a sample variance of   $20 \mu K$ in $\Delta T$,
both for Gaussian and for $\chi^{2}$ distributions.
 Minimum
sample size for current experiments are shown in Table 2.
Sky coverages of ARGO (de Bernardis et al. 1994)
 and Saskatoon (Wollack et al. 1994) is already large enough so that
the sample variance is subdominant in each experiment if the sky distribution
is Gaussian. However, substantial sky coverage  is needed
if the sky distribution is highly non-Gaussian, especially for those
experiments (MAX $\&$ MSAM) that samples around the Doppler peak
of the radiation power spectra, because the theoretical predictions of
the temperature is much higher.

If the CMB sky distribution is highly non-Gaussian on degree scales,
the signature of non-Gaussianity may well
 be detected in the large angular scale
CMB maps (Smoot et al. 1992; Ganga et al. 1993). We will use the skewness,
\begin{equation}
\mu_{3} = {\langle (\Delta T)^{3}\rangle
 \over
\langle (\Delta T)^{2}\rangle^{3/2}}
\end{equation}
 to characterize the lowest order deviation
of CMB distribution from a Gaussian.
 The skewness for the  $\chi^{2}_{n}$ distribution is $\mu_{3} = \sqrt{8/n}$.
If  the sky distribution is a $\chi^{2}_{n}$
 field on an angular scale $\theta_{0}\sim 1^{\circ}$, after smoothing
 with beam size $\theta_{s}$, the distribution is
$\chi^{2}_{n^{\prime}}$, with $n^{\prime} = n ({\theta_{s}/ \theta_{0}})^{2}$.
  For COBE, $\theta_{s} = 7^{\circ}$ and $ n^{\prime} \approx 50 n$, and
for FIRS, $\theta_{s} = 3.8^{\circ}$ and $n^{\prime} \approx 14 n$.
Thus for models with $n \gtwid 4$, the skewness  of temperature fluctuation
on COBE and FIRS scales is at least
$ \mu_{3} \gtwid 0.1 \ \ {\rm for \ \ COBE} {\rm and} \ \
\mu_{3}  \gtwid 0.30 \ \ {\rm for \ \ FIRS}.$
The cosmic variance of the skewness $\mu_{3}$ is  $\mu_{3} \sim 0.18$
 for  large
angular scale experiments like COBE or FIRS (Srednicki 1993).
 For COBE, the cosmic variance is larger
than
 the non-Gaussian signals, thus it will be
 impossible for the non-Gaussian signal
to be  detected. However, for FIRS, the smooth non-Gaussian field
will still stand above the cosmic variance.

 Finally, we shall comment on  reionization.
 Early reionization is likely if the density
fluctuations are non-Gaussian and  objects can form at an earlier epoch.
 If the universe
was reionized at an early epoch ($z \sim 100$), then  the
degree scale temperature anisotropies are dramatically reduced. In this case
the sample variance will also be reduced dramatically. As an example, in Table
2  we will also list sample variance (for different experiments)
 for a re-ionized CDM model with optical depth
$\tau =1$ (Kamionkowski et al. 1994). We conclude that we cannot reconcile
the high detection of MAX/GUM experiment with theoretical predictions
for any  $\chi^{2}$ distribution. If the MAX/GUM result is confirmed,
models with early reionization will be ruled out regardless of the
statistics of CMB at half-degree scales.

We want to thank  D. Scott and B. Moore for their helpful comments.
This work was supported by a grant from the NSF.

\newpage

\def\ref{\par\noindent\hangindent=2pc \hangafter=1 }
\centerline{REFERENCES}

\bigskip

\ref
de Bernardis, P. et al. 1994, ApJ, in press (ARGO)

\ref
Cheng, E.S., et al. 1994, ApJ, in press (MSAM)

\ref
Coulson, D., Ferreira, P., Graham, P., \& Turok, N. 1994,
Nature, 368, 27

\ref
Ganga, K., Cheng, E.S., Meyer, S. \& Page, L. 1993,  ApJ, 410, L57
(FIRS)

\ref
Gunderson, J.O., et al. 1993, ApJ, 413, L1 (MAX/GUM)

\ref
Luo, X. 1994, Phys. Rev. D, in press

\ref
Meinhold, P.R., et al. 1993, ApJ, 409, L1 (MAX/MuP)

\ref
Schuster, J. et al. 1993, ApJ, 412, L47 (SP91)

\ref
Scott, D, Srednicki, M, \& White, M. 1994, ApJ, in press

\ref
Smoot, G.F. et al. 1992, ApJ, 396, L1 (COBE)

\ref
Srednicki, M. 1993, ApJ, 416, L1

\ref
Kamionkowski, M., Spergel, D.N., \& Sugiyama, N. 1994, CfPA-94-01

\ref
Tucker, G.S., Griffin, G.S., Nguyen, H.T., \& Peterson, J.B. 1993,
ApJ, 419, L45 (White Dish)

\ref
Turok, N, \& Spergel, D.N. 1991, Phys. Rev. Lett., 66, 3039

\ref
Walker, T.P., Steigman,  G., Schramm, D.N., Olive, K.A. \& Kang, H.S.
1991, ApJ, 376, 51

\ref
White, M, Scott, D., \& Silk, J. 1994, Ann. Rev. Astro. Astrophys. in press

\ref
Wollack, E.J., Jarosik, N.C., Netterfield, C.B., Page, L.A. \&
Wilkinson, D. 1993, ApJ, 419, L49 (Saskatoon)

\vspace{0.3 in}

\newpage

\vspace {0.2 in}

\centerline{Table Caption}

\vspace {0.3in}

\noindent
Table 1: A brief summary of the experimental situation at degree scales.

\noindent
Table 2: The sample variance $\sqrt{\sigma_{sam}}$ of {\it rms} temperature
anisotropies
in various experiments, in units of $\mu K$,
for Gaussian and $\chi^{2}_{n}$ sky distributions.
The theoretical predictions for various experiments are given for a CDM
dominated universe with $\Omega=1, \Omega_{b} = 0.06$, $h=0.5$ and a
COBE normalized Harrison-Zel$'$dovich spectrum.  The numbers in bold
face are for standard recombination and those in brackets are for
a fully reionized universe with optical depth $\tau =1$.
\newpage

\vskip 0.1 in
\centerline{Table 1}
\vskip 0.2 in
\begin{tabular} {|c|c|c| c|c|  c|}
\hline
Experiment & $\theta_{FWHM}$& $\theta_{chop}$& $\theta_{c}$ &sample size&
$\Delta T$($\mu K$)($\pm 2\sigma$ ) \\ \hline
ARGO &$52^{\prime}$&$1.8^{\circ}$& $30^{\prime}$&63 & 41 - 71  \\ \hline
MAX (GUM)&$0.5^{\circ}$&$1.3^{\circ}$&$25^{\prime}$&165& 85 - 162  \\ \hline
MAX (MuP)& - & - & -&21 & $< 68$   \\ \hline
MSAM (single)&$0.5^{\circ}$&$40^{\prime}$&$.5^{\circ}$&14& 16 - 60  \\ \hline
MSAM (double)& - & - &  $.3^{\circ}$&- &30 - 85 \\ \hline
Saskatoon & $1.44^{\circ}$ &$2.45^{\circ}$&$1.44^{\circ}$ & 48& 28 - 57 ($\pm
1\sigma$) \\ \hline
SP91 &$1.5^{\circ}$ & $2.95^{\circ}$ & $1.5^{\circ}$& 13 & 13 - 45 \\ \hline
White dish & $12^{\prime}$ & $23.6^{\prime} $&$0.15^{\circ}$& 5 & $< 63$ \\
\hline
\end{tabular}
\vskip 0.3 in

\centerline{Table 2}
\vskip 0.2 in
\begin{tabular} {|c|c| c|c|  c|c|c|}
\hline
Experiment & $(\Delta T)_{th} (\mu K)$&$N_{e}$ & $N_{min}^{G}$&
$N_{min}^{\chi^{2}}$& Gaussian & $\chi_{n=1}^{2}$  \\ \hline
ARGO &{\bf 44}(27)& 63& {\bf 47}(7) & {\bf{289}}(49)&{\bf 19}(11) &{\bf 31}(18)
\\ \hline
MAX (GUM)& {\bf 71}(38)&14&{\bf 318}(26)&{\bf 2226}(182)&{\bf 44}(23)&{\bf
71}(38)  \\ \hline
MAX (MuP)& - & 12  &- &-& {\bf 45}(24)&{\bf 74}(39)  \\ \hline
MSAM (single)&{\bf 76}(44)& 8 &{\bf 417}(47)&{\bf 2919}(329)&{\bf 54}(31) &{\bf
88}(50) \\ \hline
MSAM (double)& {\bf 46}(22) &13&{\bf 56}(3) &{\bf 392}(21) &{\bf 29}(14)&{\bf
47}(23)  \\ \hline
Saskatoon & {\bf 38}(27) &27 & {\bf 26}(7)&{\bf 182}(49)&{\bf 20}(14) &{\bf
33}(23)   \\ \hline
SP91 &{\bf 38}(27)  & 6& {\bf 26}(7) &{\bf 182}(49)& {\bf 29}(21) &{\bf 47}(33)
 \\ \hline
White dish &{\bf 41}(19) & 5& {\bf 35}(2) &{\bf 245}(14)&{\bf 33}(15) &{\bf
54}(25)   \\ \hline
\end{tabular}

\end{document}